\documentstyle[aps,prl,preprint]{revtex}
\begin{document}
\draft
\title{Domain Coarsening in Systems Far from Equilibrium}
\author{M.C. Cross and D.I. Meiron}
\address{Condensed Matter Physics and Applied Mathematics\\
Caltech, Pasadena CA 91125}
\date{\today }
\maketitle

\begin{abstract}
The growth of domains of stripes evolving from random initial conditions is
studied in numerical simulations of models of systems far from equilibrium
such as Rayleigh-Benard convection. The size of the domains deduced from the
inverse width of the Fourier spectrum is found to scale as $t^{1/5}$ for
both potential and nonpotential models. The morphology of the domains and
the defect structures are however quite different in the two cases, and
evidence is presented for a second length scale in the nonpotential case
growing as $t^{1/2}$.
\end{abstract}


When a system is quenched through a transition from a disordered to an
ordered state small domains of the different symmetry manifestations of the
ordered phase initially form. These then grow to give larger ordered
regions, asymptotically approaching the ideal of very large ordered regions.
There has been a great deal of study of this {\em coarsening} process in
systems where the different states are equilibrium thermodynamic phases at
finite temperature. In this paper we extend this question to systems in
which the ``ordered'' state is produced by a pattern forming instability in
a system far from equilibrium (the canonical example being Rayleigh-Benard
convection).

We investigate the formation of a ``stripe'' state in two dimensions, with
rotational invariance in the plane. Rayleigh-B\'{e}nard convection is such a
system, with the stripes corresponding to the familiar convection rolls.
After a quench into the ordered region, given by stepping the Rayleigh
number (the dimensionless temperature difference across the depth of the
cell driving the convection), regions of differently oriented stripes grow
from random initial fluctuations as the dynamics rapidly drives the system
locally to a state with the characteristic length scale of the stripes. The
questions that arise are how does the length scale over which the stripes
are ordered grow, and what are the scaling properties, for example
characterized by the structure factor. In addition, because of the stripe
nature of the ordered state, there is the question of what is the large
scale morphology, for example is it best described as domains of rather
straight stripes with sharp domain boundaries between them (i.e. a pattern
of grains), or are the stripes curved on the large length scale. (This same
question would arise in the quenching into an equilibrium stripe phase such
as a two dimensional smectic.) For the system far from equilibrium there is
also the question of ``wavevector selection'' i.e. what is the long time
asymptotic wavelength of the stripe pattern. Note that for a thermodynamic
equilibrium phase the asymptotic wavevector will simply be the one that
minimizes the free energy. For a stripe state far from equilibrium there is
no corresponding argument, and indeed the question of wavevector selection
has aroused considerable interest.

In this paper we study the formation of the stripe phase in numerical
simulations of equations that model Rayleigh-B\'{e}nard convection. These
equations are based on the Swift-Hohenberg equation that was introduced \cite
{SH} to look at the effect of fluctuations on the transition to the
convective roll state. It is an equation for a real order parameter $\psi (%
\vec{r},t)$ that is a function of the horizontal coordinates $\vec{r}=(x,y)$
and time $t$. In a uniform convective state of straight parallel rolls $\psi
$ takes the form
\begin{equation}
\psi \propto \cos (\vec{q}\cdot \vec{r}+\phi )+harmonics  \label{OP}
\end{equation}
with $\vec{q}$ the wavevector of the stripes and $\phi $ the phase variable,
and gives the horizontal variation of the pattern that is involved in
questions of pattern formation. Swift and Hohenberg were interested in
universal aspects of the transition, and so wrote down the simplest
dynamical equation consistent with the symmetries and the existence of a
stripe state:
\begin{equation}
\dot{\psi}=\epsilon \psi -(\nabla ^2+1)^2\psi -\psi ^3+\eta (\vec{r},t)
\label{SH}
\end{equation}
with the dot denoting a time derivative and $\nabla ^2$ the two dimensional
Laplacian. Here $\epsilon $ is the control parameter, depending linearly on
the temperature difference driving the convection, with the transition to
stripes occurring for $\epsilon >0,$ and $\eta $ is a noise term that was
introduced to investigate the effect of thermal fluctuations on the
transition. Equation \ref{SH} also describes a near equilibrium system,
since the dynamics are ``potential'' i.e. follow the descent of a potential
functional, which would be the free energy for the near equilibrium system.
Eq. \ref{SH} has since been much used as a useful qualitative model of
features of pattern forming systems that may not be quantitatively
universal, since it incorporates the three important features of such
systems, namely growth of the disturbance, non-linear saturation, and
dispersion. However the potential aspect of the dynamics is not appropriate
for systems far from equilibrium, and so various modifications have been
proposed to account for this aspect. In particular Greenside and Cross \cite
{GC} suggested a modification of the non-linear term to yield
\begin{equation}
\dot{\psi}=\epsilon \psi -(\nabla ^2+1)^2\psi +3(\vec{\nabla}\psi )^2\nabla
^2\psi \qquad .  \label{NPP}
\end{equation}
As well as being non-potential, this equation also gives a better
representation of the stability of the stripe phase as the wavevector and
control parameter are varied (the ``stability balloon'') for convection \cite
{GC}. A further model is given by incorporating the effects of mean drift.
In the convection system a mean flow (averaged over the depth of the cell),
slowly varying with the horizontal coordinates, is an important additional
degree of freedom that qualitatively changes the physics \cite{SZ}, \cite{C}%
. This leads to the equations
\begin{eqnarray}
\dot{\psi}+\vec{U}.\vec{\nabla}\psi  &=&\epsilon \psi -(\nabla ^2+1)^2\psi
+NL  \nonumber \\
\vec{\nabla}\times \vec{U} &=&\Omega \hat{z}  \label{SHV} \\
\tau _v\dot{\Omega}-\sigma (\nabla ^2-c^2)\Omega  &=&g_m\hat{z}\cdot \vec{%
\nabla}(\nabla ^2\psi )\times \vec{\nabla}\psi   \nonumber
\end{eqnarray}
where $\vec{U}(x,y,t)$ is the divergence-free horizontal velocity that
advects the field $\psi $ in the first equation (with the symbol $NL$
referring to either of the non-linearity choices). The velocity $\vec{U}$ is
defined in terms of the vertical vorticity $\Omega $ which is in turn driven
by distortions of the stripe pattern through the third equation. Here $g_m$
gives the coupling between the mean flow and the stripes, and increases as
the Prandtl number (the ratio of viscous to thermal diffusivities)
decreases. The parameters $\tau _v,$ $\sigma $ and $c$ may be chosen to
match to the fluid system: we use $\tau _v=1,\;\sigma =1,$ and $c^2=2$. We
will present results for $\epsilon =0.5$ for eqs. \ref{NPP} and \ref{SHV},
and $\epsilon =0.25$ in eq. \ref{SH} since the dynamics appears to freeze at
long times in this model for the higher value.

Coarsening in the Swift-Hohenberg equation has been studied by Elder et al.%
\cite{EVG}. In the presence of the noise term (corresponding to a finite
temperature thermodynamic system) their numerics yielded a length scale
increasing with time as $t^{1/4}$. In the absence of noise they found a
slower growth, consistent with a $t^{1/5}$ scaling. Elder et al. were
surprised by these results. They expected the long time dynamics of the
system to be governed by the ``phase diffusion equation'', an equation for
the slow space and time variation of the phase $\phi $ introduced in eq. \ref
{OP}:
\begin{equation}
\dot{\phi}=D_{\Vert }\partial _{\Vert }^2\phi +D_{\bot }\partial _{\bot
}^2\phi   \label{Phase}
\end{equation}
where $\Vert $ and $\bot $ refer to directions parallel and perpendicular to
the local wavevector and $D_{\Vert }(q)$ and $D_{\perp }(q)$ are diffusion
constants that depend on the local wavenumber $q=\left| \vec{\nabla}\phi
\right| $. Simple power counting leads to a length scale growing
asymptotically as $\xi \sim t^{1/2}$ \cite{EVG}. Earlier however, Cross and
Newell \cite{CN} had proposed that on long time scales the local wavenumber
should tune itself to a value $q_f$ for which $D_{\perp }=0$, so that the
second term in eq. \ref{Phase} drops out. In addition the first term in eq.
\ref{Phase} is also zero since $\partial _{\Vert }\phi =q$ is then constant.
Cross and Newell proposed that higher order gradient terms in the phase
equation would control the dynamics. In the original paper they suggested a
scaling $\xi \sim t^{1/3}$, but more recently Cross and Hohenberg \cite{CH}
suggested $t^{1/4}$ as the correct result of this analysis, closer to the
results of the numerics. The tuning of the wavenumber to $q_f$ follows
directly for potential systems, since this wavenumber is also the one that
minimizes the potential. However Cross and Newell suggested that this was
also valid more generally, since in the absence of coupling to mean drift,
focus singularities also relax the wavenumber to this value. (A focus
singularity is the center of curvature of axisymmetric stripes or a sector
of such stripes, and permits the disappearance or nucleation of stripes
driven by curvature effects until $D_{\perp }\rightarrow 0$.) This tuning
does not survive the coupling to mean flow effects. Thus at the outset of
this work we expected to find a $t^{1/4}$ or $t^{1/5}$ scaling for the
nonpotential case eq. \ref{NPP} with a concomitant approach of the mean
wavenumber to $q_f$, but a different scaling, perhaps $t^{1/2}$, with the
addition of mean flow.

We now summarize our results. We find a slow evolution of the characteristic
length scale $\xi _q$, defined from the width of the wavevector distribution
in Fourier space, consistent with $\xi _q\sim t^{1/5}$ for {\em all three}
cases studied: potential, non-potential, and with the inclusion of mean flow
(also non-potential). For the non-potential cases, a second length scale $%
\xi _r$ defined from the correlations of the orientation of the stripes in
real space (a less accurate calculation) appears to show a different scaling
consistent with $\xi _r\sim t^{1/2}$. In the non-potential cases the
asymptotic wavenumber does {\em not} approach the value $q_f$ at which $%
D_{\perp }$ is zero. Rather it seems to approach closely the wavenumber at
which isolated dislocations are stationary. The morphology of the pattern
appears quite different in the potential and nonpotential cases (see Fig.
\ref{fig1}). For the potential equation the pattern (panel a) may be
described as largely consisting of domains of straight stripes with sharp
boundaries between the domains, although there are also some regions where a
smoother variation of the stripes is seen. Two kinds of domain walls may be
identified: lines where one set of stripes ends and a second set starts,
visible in panel b; or lines where there is a sharp kink in the stripe
orientation, but a smaller perturbation of the amplitude so that there is no
signature in panel b. For the non-potential equation stripes smoothly curved
over the characteristic scale are evident (panel c), and isolated
dislocation defects are more apparent (see also panel d).

The variation of the length scale $\xi _q$ with time is shown in Fig. \ref
{fig2}, showing results from our longest runs (times to 7400) and largest
systems (around 1000 in size). These results were produced from random
initial conditions (independent random numbers on each mesh point) with a
time step of 0.2, after an initial transient of time length 1 integrated
with time step .01 to allow the large wave vector components of the initial
condition to decay. The numerical scheme was a pseudospectral scheme using $%
1024\times 1024$ Fourier modes with second order accurate time stepping
described previously \cite{CTM}, with periodic boundary conditions. Each
integration step took $0.5$ - $1$ second (depending on the complexity of the
nonlinear term in the equation) on one processor of a CRAY C90.

Figure \ref{fig2} shows the time variation of the ``width'' $\delta q$ of
the structure factor $S(q,t)=<\psi (\vec{q},t)\psi (-\vec{q},t)>$ in Fourier
space, with both axes on logarithmic scales. The average $<>$ is over angles
of the wavevector $\vec{q}$, and the data represents a single run, although
other runs were done with consistent results. We extract the width $\delta q$
from a Lorenzian squared fit
\begin{equation}
S(q)=\left( \frac a{(q^2-b)^2+c^2}\right) ^2
\end{equation}
with $\delta q$ defined as the half width at half height $\delta q=0.322c/%
\sqrt{b}$. With this fit function the values extracted do not significantly
depend on the range over which the function is fitted: we have used a range $%
\pm 4\delta q$. For both the potential Swift-Hohenberg model (as seen
earlier by Elders et al. \cite{EVG}) and the non-potential model the long
time variation is well fit by a power law $\delta q\propto t^{-1/5}$ over
two or more decades in time \cite{log}. This same scaling was found
including the mean flow: we used vorticity coupling constants $g_m=10$ and $%
g_m=20$ with $\tau _v=1$, $c^2=2$, $\sigma =1$, in a smaller system size of
about $536$ with a $512\times 512$ mesh because of the increased numerical
complexity, and to shorter times (about $4000$) to eliminate finite size
saturation effects. The results remain consistent with a power law scaling
of around $\frac 15$, and are clearly not consistent with a $t^{1/2}$
scaling.

We can also wonder whether the patterns are characterized by a single (long)
length scale. To investigate this question we have calculated the stripe
orientation field correlation function. The Fourier space filtering method
to extract the local orientation $\theta (\vec{r},t)$ of the stripes has
been described previously \cite{CTM}. We then calculate $C_2(|\vec{r}-\vec{r}%
^{\prime }|,t)=<e^{i2(\theta (\vec{r},t)-\theta (\vec{r}^{\prime },t))}>$
averaging over the spatial coordinates $\vec{r}$ and $\vec{r}^{\prime }$ for
fixed $r=\left| \vec{r}-\vec{r}^{\prime }\right| $ for each time $t$. As
expected the correlations decay with increasing separation $r$, and from
this decay we can extract a characteristic length $\xi (t)$. (We have chosen
to use simply the half width at half height of $C_2$, since the quality of
the data and the limited range of $r$ probed due to the finite size of the
system do not warrant a functional fit to the decay.) Interestingly, for the
potential Swift-Hohenberg model we find that the variation of $\xi $ is
roughly consistent with $\delta q^{-1}$ from the width of the structure
factor (the fit shown in the inset to Fig. \ref{fig2} over a relatively
small range of times gives $\xi \sim t^{0.24}$). On the other hand for the
nonpotential model the variation is close to $\xi \sim t^{1/2}$ rather than
the $t^{1/5}$ for $\delta q^{-1}$. Inspecting the domain morphology from
Fig. \ref{fig1}, or better from a plot of the orientation field $\theta (%
\vec{r})$, suggests that in the nonpotential case there may be stronger
correlations along the stripes, than perpendicular to the stripes. We might
expect the latter to have a stronger influence on the range of wavenumbers
in the pattern that determines $\delta q$.

An important question that has aroused much interest in nonequilibrium
stripe states is the question of wavevector selection i.e. is there a
preferred wavevector for each set of control parameters that is selected in
patterns under a wide range of situations such as different geometries or
initial conditions. One popular hypothesis in the literature has been the
``maximum growth rate'' idea, namely that the wavevector selected is the one
that has the maximum growth rate in the linear (small amplitude) regime.
More recently the importance of the nonlinearity of the system has been
studied, mainly in two classes of situations: the evolution from random
initial conditions in a {\em one dimensional} geometry; and geometrically
simple situations in one or two dimensions, for example concentric stripes,
or patterns with one or two defect structures such as dislocations or grain
boundaries (see ref. \cite{CH} for a review). The present situation combines
aspects of both of these: we start from random initial conditions, but also
the evolution leads to states with defects that allow wavenumber relaxation.
In the potential case all results have shown that the wavevector that
minimizes the potential is selected in all cases. For the Swift-Hohenberg
model at $\epsilon =0.25$ this is very close to $q=1$, and indeed we see,
Fig. \ref{fig4}, that the mean wavenumber is very close to this value over
the whole time period. For the non-potential cases the evolution is much
more interesting. Although a value $q\simeq 1$ is produced by the early
evolution where the biharmonic operator tends to filter out other
wavenumbers in the linear evolution (i.e. consistent with the maximum growth
rate idea), at later times there is a trend to smaller wavevectors, and then
a slow increase to a long time asymptotic value away from unity. Empirically
we find that the wavenumber at long times approaches a value that is very
close to the {\em wavenumber }$q_d${\em \ at which isolated dislocation
defects are stationary} i.e. have zero climb velocity. (These values are
obtained by separate runs measuring the climb velocity of a defect pair in
otherwise straight stripes at various wavenumbers, and the wavenumber for
zero climb velocity is found by interpolation.) The relevance of the
dislocation selected wavenumber is confirmed by the agreement of the
asymptotic wavenumber with $q_d$ for the two different values of $g_m$, as
shown in Fig. \ref{fig4}.

We do not have a good theoretical understanding of these results. In the
potential case, a dimensional analysis of the phase equation might be argued
to lead to a $t^{1/4}$ scaling since the diffusion constant tends to zero as
the wavenumber approaches its long time asymptote. Although this result was
found by Elders et al. \cite{EVG} in the case with added noise, both they
and we find a slower dependence in the absence of noise, although the
expectation \cite{B}, \cite{RB} is that noise should be irrelevant to the
long time, large scale dynamics. Recent work \cite{RB} has suggested the
importance of defects in the coarsening process leading to corrections to
this naive result. The defect structure of stripe phases is complicated, and
understanding the role of the various defects in the coarsening process as
well as an extension of these ideas to the non-potential case remain
challenges for the future.

\begin{figure}[tbp]
\caption{ Morphology of the stripe pattern in (a) potential and (c)
non-potential models of convection during the coarsening process, with black
and white denoting positive and negative values of the field $\psi $. The
pictures show one quarter of the acutal system used. Panels (b) and (d) show
the corresponding defect structure visualized as the regions where the
amplitude of the stripe pattern (calculated using Fourier filtering methods)
falls below 75\% of the maximum value.}
\label{fig1}
\end{figure}

\begin{figure}[tbp]
\caption{Plot of the logarithm of the inverse length scale as a function of
log time for the potential (squares) and nonpotential (crosses) models. The
main figure shows the width of the structure factor; the inset shows the
inverse of the half-width of the decay of the orientation field correlation
function in real space. }
\label{fig2}
\end{figure}

\begin{figure}[tbp]
\caption{Evolution of the mean wavenumber with log time for the potential
Swift-Hohenberg model (triangles), the nonpotential model (squares) and the
nonpotential model with added mean flow (circles). Also shown are some
characteristic wavenumbers {\em for the nonpotential model}: E -- the
wavenumber at which the stripes become unstable to the longitudinal
(Eckhaus) instability; $q_f$ -- the wavenumber selected by focus defects at
which $D_{\perp }$ goes to zero in the absence of mean flow; and $q_d$ --
the wavenumber at which the climb velocity goes to zero, shown for the two
values of the vorticity coupling used in the plot. }
\label{fig4}
\end{figure}

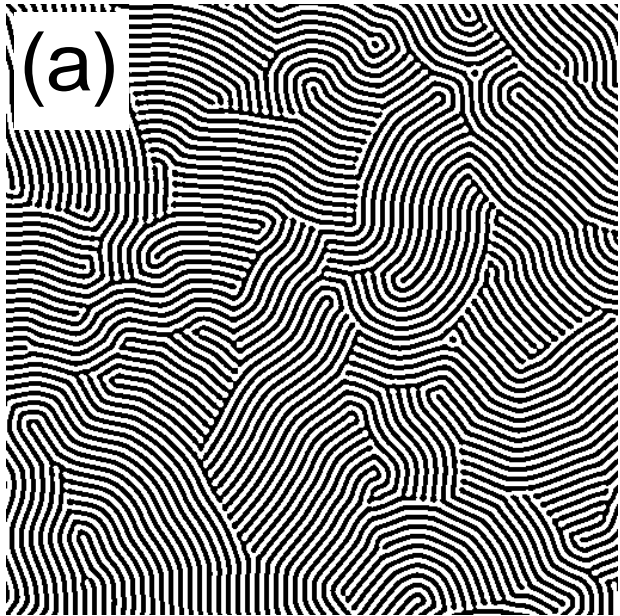
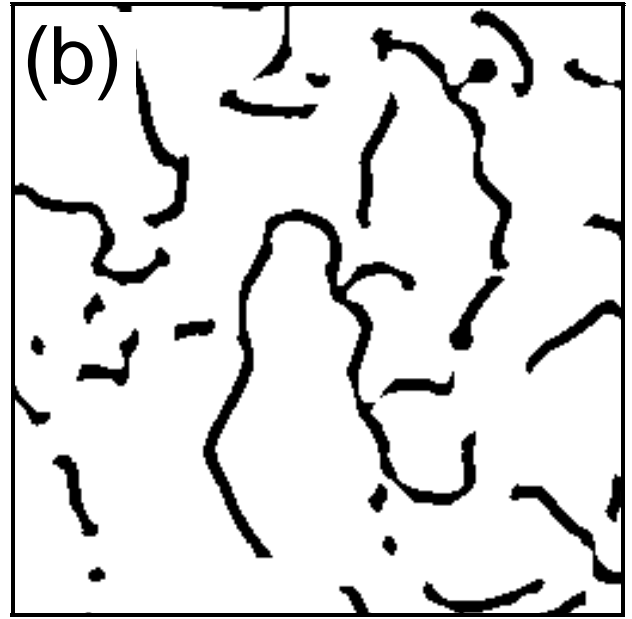
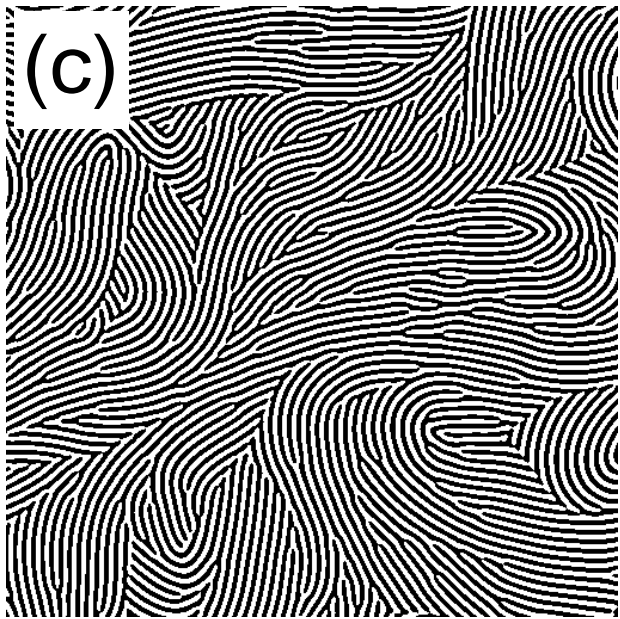
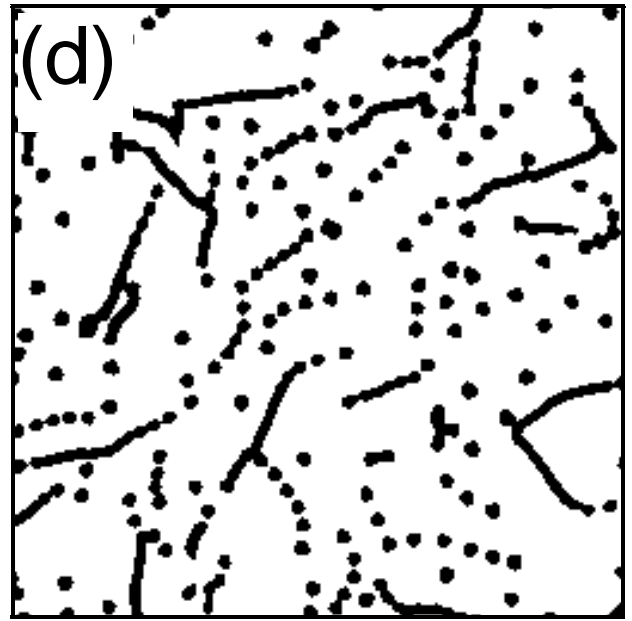
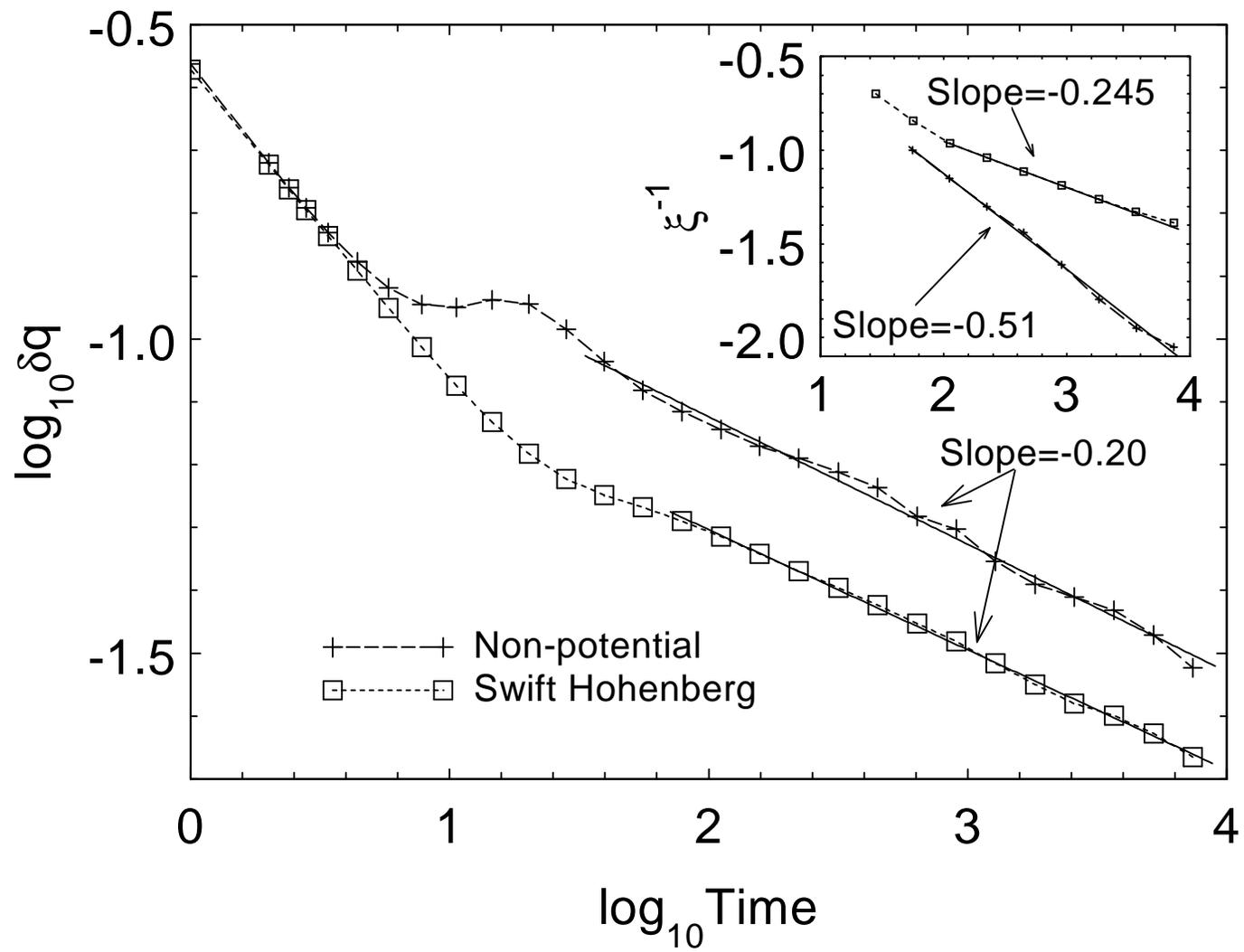
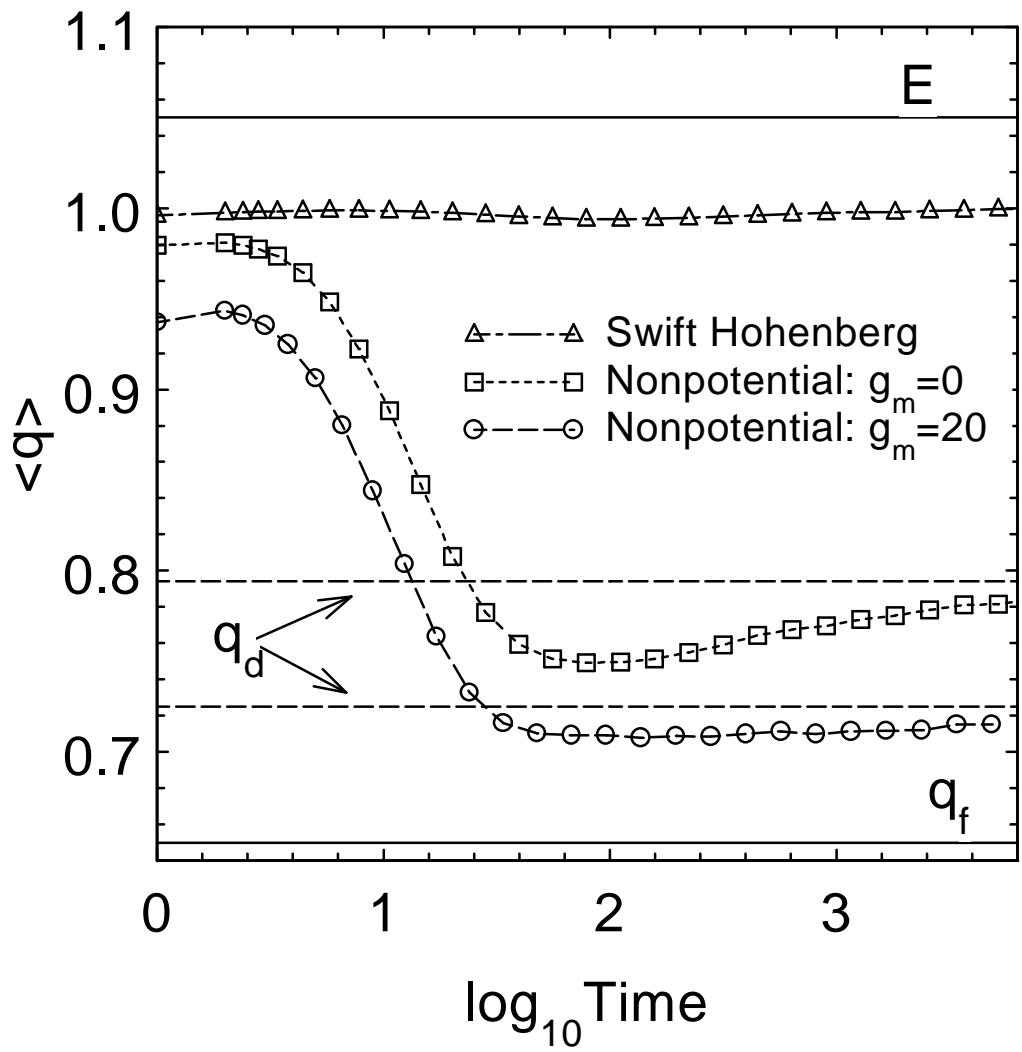
\end{document}